# Imaging Ferroelectric Domains via Charge Gradient Microscopy Enhanced by Principal Component Analysis

Ehsan Nasr Esfahani[1,2], Xiaoyan Liu[3,*], and Jiangyu Li[1,2,*]

[1]    Department of Mechanical Engineering, University of Washington, Seattle, WA 98195, USA

[2]    Shenzhen Key Laboratory of Nanobiomechanics, Shenzhen Institutes of Advanced Technology, Chinese Academy of Sciences, Shenzhen 518055, Guangdong, China

[3]    College of Metallurgy and Material engineering, Chongqing Key Laboratory of Nano/Micro Composites and Devices, Chongqing University of Science &Technology, Chongqing, China

## Abstract

Local domain structures of ferroelectrics have been studied extensively using various modes of scanning probes at the nanoscale, including piezoresponse force microscopy (PFM) and Kelvin probe force microscopy (KPFM), though none of these techniques measure the polarization directly, and the fast formation kinetics of domains and screening charges cannot be captured by these quasi-static measurements. In this study, we used charge gradient microscopy (CGM) to image ferroelectric domains of lithium niobate based on current measured during fast scanning, and applied principal component analysis (PCA) to enhance the signal-to-noise ratio of noisy raw data. We found that the CGM signal increases linearly with the scan speed while decreases with the temperature under power-law, consistent with proposed imaging mechanisms of scraping and refilling of surface charges within domains, and polarization change across domain wall. We then, based on CGM mappings, estimated the spontaneous polarization and the density of surface charges with order of magnitude agreement with literature data. The study demonstrates that PCA is a powerful method in imaging analysis of scanning probe microscopy (SPM), with which quantitative analysis of noisy raw data becomes possible.

---

* To whom the correspondence should be addressed to; email: jjli@uw.edu or liushoen@163.com

**Introduction**

Lithium niobate (LiNbO$_3$) is a uniaxial ferroelectric crystal with large spontaneous polarization ($Ps = 80 \pm 5 \frac{\mu C}{cm^2}$) aligned along the crystallographic $Z$-axis [1, 2]. Because of its versatile ferroelectric properties, LiNbO$_3$ has been used in a wide ranges of applications, including electro-optics [3], nonlinear optics [4, 5], ferroelectric data storage [6, 7], and microelectromechanical devices [8]. Most of these applications involve ferroelectric domains with 180° domain walls created by applying an external electric field to produce antiparallel (180°) domains with $+P_s$ and $-P_s$ polarizations, and the large spontaneous polarization of LiNbO$_3$ often results in screening charges from the surrounding environment, so that bound charges on the crystal surface can be compensated, and the electrostatic energy can be minimized [9].

Local domain structure and dynamics of domain walls in LiNbO$_3$ have been studied extensively using various modes of scanning probe microscopy (SPM) techniques at the nanoscale, including piezoresponse force microscopy (PFM) [10-16], Kelvin probe force microscopy (KPFM) and time-resolved Kelvin probe force microscopy (tr-KPFM) [17, 18], and electrostatic force microscopy (EFM) [19-22]. Because of experimental complications, all these SPM techniques are usually performed with a slow or moderate scan speeds, less than 10 Hz/line over 10 $\mu m$ length, with which the pixel time of imaging is much longer than the time scale of the ferroelectric dynamics. Therefore, the fast formation kinetics of domains and domain walls and the evolution of the screening charges cannot be captured by such quasi-static measurements. Furthermore, none of these techniques measure the polarization directly, and thus the data interpretation is often challenging. For example, PFM images ferroelectric domains through piezoelectric strain, and it is now well known that multiple electromechanical mechanisms contribute to the piezoresponse signal measured by PFM [23-27].

Recently, charge gradient microscopy (CGM) has been developed to study the ferroelectric domains and characterize their surface charges [21, 28, 29]. It operates by mechanically scraping the screening charges on the surface using a conductive scanning probe, and collecting the resultant current using conductive atomic force microscopy (cAFM). Note that these scraped charges can be compensated and refilled by the conductive probe, and since the probe is virtually grounded under cAFM, the refilled charges can be measured as a current signal, making imaging possible. Because of the fast-moving probe, the refilling process is rapid compared to other mechanisms, enabling the study of dynamic process of domains and domain walls. More importantly, the current measured can be directly related to polarization, unlike other SPM techniques discussed earlier.

In this paper, we study ferroelectric domain structure of $LiNbO_3$ using CGM, and investigate the effects of scan speed and temperature on CGM signals. We rely on principal component analysis (PCA) to reduce the dimensionality of the data and enhance their signal-to-noise ratio, and demonstrate that CGM signal increases linearly with the scan speed while decreases with the temperature under power-law. The study suggests that current measured under CGM arise from the scraped and refilled surface charges within a domain, and from polarization change experienced by the scanning probe across a domain wall, enabling us to estimate the spontaneous polarization and surface charge density of $LiNbO_3$.

**Methods**

**_Experimental methods_**

We used periodically poled $LiNbO_3$ (PPLN, Asylum Research, USA) sample in our study, which is mounted on a metal puke and has dimensions of $3\ mm\ \times\ 3\ mm\ \times 0.5\ mm$, with the poled domains approximately $10\ \mu m$ in width. All the experiments were carried out on Asylum Research atomic force microscopes (Cypher and MFP-3D) using a conductive diamond probe (CDT-NCHR-10, Nanosensors, Inc.) with a spring constant around $80 \frac{N}{m}$.

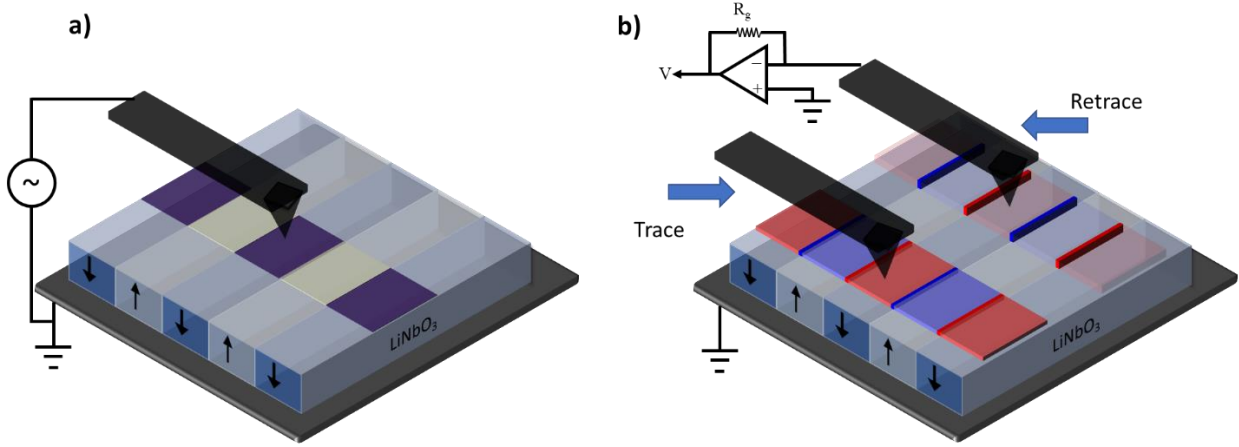

**Fig. 1** Schematics of (a) PFM with phase map overlaid partially; and (b) CGM with trace and retrace current mappings overlaid.

As a benchmark, PFM scans were carried out first to map the ferroelectric domain structure of PPLN, as schematically shown in Fig. 1(a). An AC voltage of 7 V with a single frequency of 135 kHz was used to excite the piezoresponse, which is far away from the contact resonant frequency of approximately 2 MHz, making it possible to identify the polarization orientation directly from the piezoresponse. The scan speed was $22 \frac{\mu m}{s}$ and the contact force was kept constant at approximately 500 nN. The CGM signals were acquired using ORCA (Asylum Research, USA), a cAFM module with a grain of $5 \times 10^8 \frac{V}{A}$, as schematically shown in Fig. 1(b). The scan speed was varied from $\sim 75 \frac{\mu m}{s}$ to $\sim 3 \frac{mm}{s}$, and the contact force in all the CGM measurements was kept constant at around 3 μN, much higher than that of PFM to ensure that surface charges were scraped effectively. In addition, we used a heating stage (PolyHeater™, Asylum Research, USA) to study the effect of temperatures on CGM. In the experiment, the temperature of the sample was slowly raised to 83℃ and kept constant. Continuous CGM imaging was then carried out while the temperature was ramped down with a rate of 4℃/min controlled via a closed-loop feedback.

### _Principal component analysis_

A series of CGM current mappings were acquired under different conditions, including different scan speeds and temperatures, and the data were post-processed by principal component analysis [30, 31] to enhance the signal-to-noise ratio using MATLAB®, as detailed in the supplementary information (SI). Under PCA, a set of $p$ current images (variables) of $m$-by-$n$ pixels (observations) is represented as:

$$I_i(\omega_j) = a_{ik}v_k(\omega_j),$$

where $I_i(\omega_j) \equiv I(x, y, \omega_j)$ is the current images data at the discrete variables $\omega_j$ (such as scan speed or temperature), $a_{ik} \equiv a_k(x, y)$ are the spatial expansion coefficients (known as PCA component images or PCA loadings), and $v_k$ are the corresponding eigenvectors (PCA coefficients) [32]. Note that the matrix of experimental data $\mathbf{I}_{ij}$ is constructed from CGM current mappings, with each column corresponding to the reshaped grid points of each current image ($i = 1, \ldots, l = m \times n$) and each row representing the experimental variables ($j = 1, \ldots, p$) for a certain pixel. The eigenvectors $v_k$ and their corresponding eigenvalues $\lambda_k$ can be found from the empirical covariance matrix or more efficiently by the singular value decomposition (SVD) of the experimental matrix:

$$\mathbf{I} = \mathbf{U}\boldsymbol{\Sigma}\mathbf{W}^T,$$

where $\mathbf{W}$ is a $p$-by-$p$ unitary matrix and each column of $\mathbf{W}$ are the eigenvectors $v_k$, $\boldsymbol{\Sigma}$ is a $l$-by-$p$ rectangular singular matrix whose singular values equal to the square root of the eigenvalues $\lambda_k$, and $\mathbf{U}$ is a $l$-by-$l$ unitary matrix. Each column of $\mathbf{T} = \mathbf{U}\boldsymbol{\Sigma}$ corresponds to the spatial expansion coefficients $a_k$, also known as PCA component images, PCA loadings, or PCA scores, and the $j^{th}$ PCA component image can be obtained by reshaping the $j^{th}$ column of $\mathbf{T}$ to a matrix of m-by-n. Note that in PCA, the eigenvectors $v_k$ are orthogonal and sorted in a way that the corresponding eigenvalues $\lambda_k$ are in descending order, i.e. the first eigenvector $v_1$ has the largest possible variance and thus, contains the most information. Dimensionally-reduced data can be reconstructed by forming a truncated $l$-by-$L$ matrix $\mathbf{I}_L = \mathbf{U}_L\boldsymbol{\Sigma}_L\mathbf{W}_L^T$ considering only the first $L$ PCA modes (the $L$ largest variances).

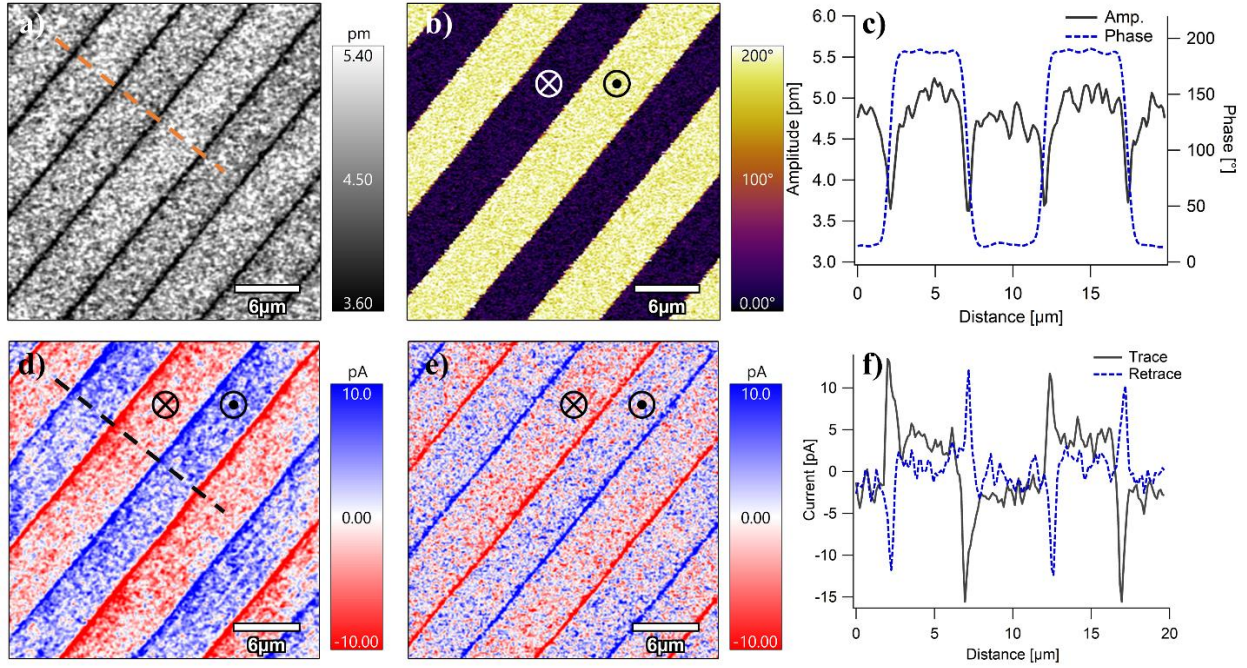

**Fig. 2** Domain pattern of PPLN mapped by (a-c) PFM and (d-f) CGM; PFM amplitude (a), phase (b), and line scans of amplitude and phase (c); and CGM current mapping during trace (d), retrace (e), and corresponding line scans (f). The PFM was acquired with probe speed of 22 $\frac{\mu m}{s}$ while the CGM acquired with probe speed of 1.5 $\frac{mm}{s}$. The upward and downward domains are shown by ⊙ and ⊗ signs, respectively.

**Results and discussions**

In order to map the domain structure and identify the polarization direction of PPLN, we first performed a slow off-resonance PFM scan, since the absolute phase of piezoresponse at the contact resonance is not well defined, and resonance-enhanced imaging at higher frequencies can change the piezoresponse due to the dynamics of cantilever or instrumental lags [33-35]. Away from the resonance, however, the polarization direction can be deduced directly from the piezoresponse phase signal measured [24]. The domain pattern of PPLN is revealed by PFM amplitude mapping in Fig. 2 (a), wherein domains of comparable amplitude are observed, separated by domain walls with much reduced response. This is evident from PFM phase mapping in Fig. 2 (b) as well, wherein 180º phase contrast is observed across domains. The variation of PFM phase and amplitude can be more clearly seen in Fig. 2(c), wherein line scans are shown. Given the positive $d_{33}$ coefficient for PPLN, the phase should be around 180° and 0° for upward-and downward-polarized domains, respectively [36], which are labelled by $\odot$ and $\otimes$ signs in Fig. 2(b).

While powerful for domain imaging, PFM does not yield any information on local polarization value. The CGM signals acquired in a separate scan but on the same area as PFM, on the other hand, reveal interesting current mappings (Fig. 2(d,e)) that match PFM well. First of all, there are current spikes when passing over domain walls, and the sign of such spikes is reversed between trace and retrace passes, and between entering and exiting a particular domain, which can be seen more clearly from the line scans in Fig. 2(f). Meanwhile, the current magnitude is reduced within domains, and the sign of the current acquired during trace and retrace are identical for the same domain, though between the adjacent domains the sign is reversed. The difference between CGM signals at domain walls and within domains can be understood from two different imaging mechanisms. As discussed in the introduction, polarization within a domain is compensated by the screen charge on the surface, and the charges are mechanically scraped by the CGM probe and then refilled, resulting in current measured. For a downward polarization, the positive surface charges are scratched away and have to be refilled from the probe, resulting in a negative current (positive charges flowing from the probe to the sample), and the sign does not depend on the scan direction. For an upward polarization, positive current (flowing from the sample to probe) is resulted. This is indeed what we observe in Fig. 2(d,e), confirming that the imaging mechanism within domains is scratching and refilling of surface charge by the virtually grounded CGM probe. When a domain wall is crossed, for example from a downward polarized domain to an upward polarized one, the net polarization difference is $+2P_s$, resulting in a positive current spike, which is reversed when going from the upward polarization to the downward one. This also explains why the current spikes reverse polarity between trace and retrace passes. These observations confirm that the imaging mechanism at domain walls is caused by the change of polarization experienced by the probe. As such, two different imaging mechanisms exist for CGM, which were observed by Hong et al. as well [21]. We also noted that Schroder et al. reported unexpected photoinduced current along domain walls in lithium niobate single crystal, though the mechanism is different there [37].

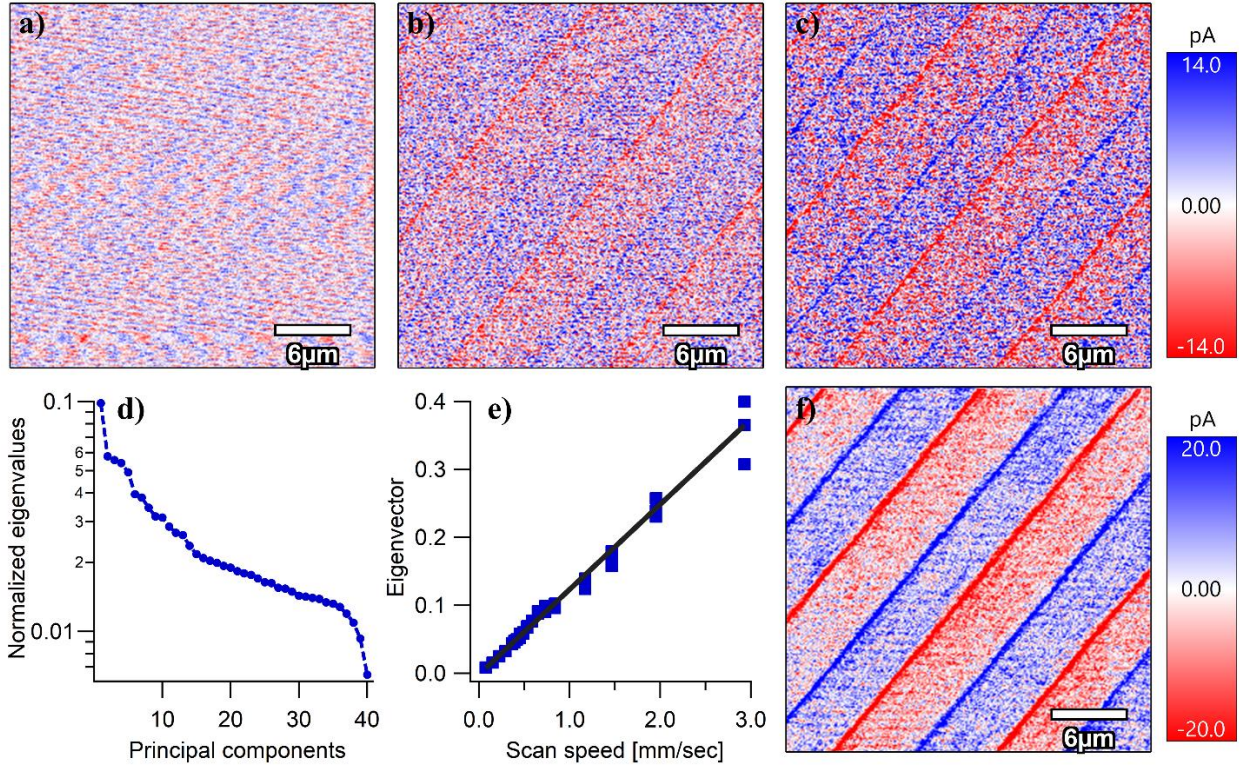

**Fig. 3** CGM under different scan speeds; (a-c) raw CGM current mappings acquired with scanning speed of 0.36 $\frac{mm}{s}$(a), 1.46 $\frac{mm}{s}$ (b), and 2.93 $\frac{mm}{s}$ (c); (d-f) PCA of scan speed-dependent images: scree plot of eigenvalues (d), the first PCA eigenvector (e), and its corresponding component image (f).

To further understand the imaging mechanism of CGM, we investigate the effect of scan speed, ranging from ~75 $\frac{\mu m}{s}$ to ~3 $\frac{mm}{s}$, on CGM signals. The raw CGM mappings acquired during the trace pass under a slow, intermediate, and fast scan speeds are presented in Fig. 3(a-c), showing in general enhanced CGM signals under higher scan speeds. However, having currents in the order of pA, the signals are rather noisy and it is difficult to draw a quantitative conclusion. Traditionally, the evolution of a set of AFM images were compared by their mean and standard deviation, which can be unreliable given the noisy CGM measurements [21]. Principal component analysis (PCA), a powerful tool for background and noise subtraction [38], was employed to analyze the data. The PCA eigenvalues drop rapidly after the first one, as shown in Fig. 3(d), suggesting that the first mode has the highest possible variance and contains the most information in this set of data. The first eigenvector $v_1(\omega)$ captures the average spectrum of data as a function of discrete set of scan speeds, as shown in Fig. 3(e), and its corresponding component image (1$^{st}$ PCA loading) shows the spatial distribution of CGM data (Fig. 3(f)). The higher component images of PCA are shown in Fig S1, confirming that the first mode is indeed sufficient to represent the data. The noise reduction is evident in the first component image, while the corresponding eigenvector exhibits a linear correlation between scan speed and the average current signal. Such a linear relationship is consistent with the imaging mechanisms proposed, as current under a constant change in charge is inversely proportional to time, and thus linear to scan speed.

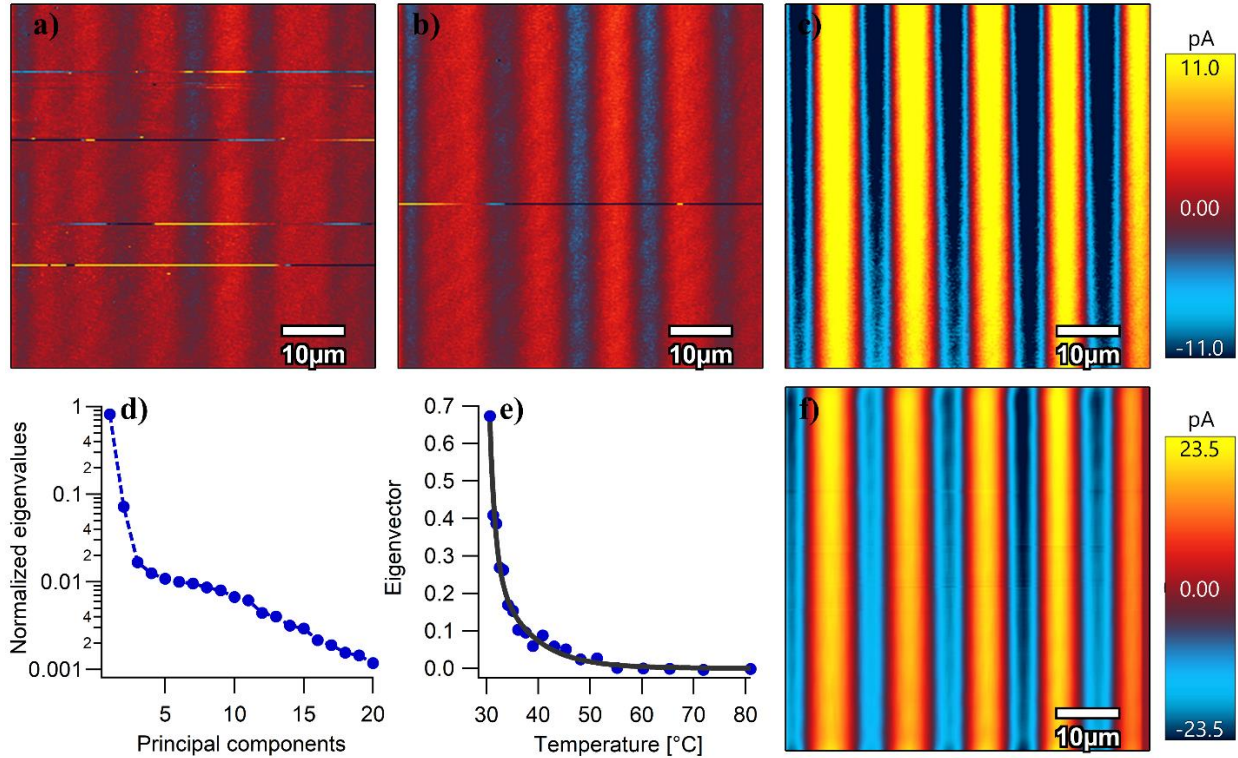

**Fig. 4** CGM under different temperatures; (a-c) raw CGM current mappings acquired under temperatures of 71.9℃ (a), 40.8℃ (b), and 30.6℃ (c); (d-f) PCA of temperature-dependent images: scree plot of eigenvalues (d), the first PCA eigenvector (e), and its corresponding component image (f). The scan speed in all 20 images was $1.2 \frac{mm}{s}$.

We then investigate the origin of current signals further by performing temperature-dependent CGM, acquired while the temperature is continuously decreasing from 83℃ to 30℃. The raw CGM current images acquired at three distinct temperatures are shown in Fig. 4(a-c), showing higher signal at the lower temperature. To see this more clearly, this set of temperature-dependent CGM data was also analyzed by PCA. The normalized eigenvalues are plotted in Fig.

4(d), and it is observed that the second and higher modes of PCA have variances orders of magnitude less than that of the first mode, as confirmed by the higher modes PCA component images in Fig. S2. As such, the first mode contains the most information and the power-law decrease with increased temperatures exhibited by the first eigenvector, as shown in Fig. 4(e), is sufficient to represent the trend of the data. The corresponding component image is shown in Fig. 4(f), again with much enhanced signal-to-noise ratio. This temperature-dependence can be understood again from the imaging mechanism of the CGM that the increased temperature reduces spontaneous polarization and removes the screening charges, and thus reduces CGM signal. Indeed, Tong et al. discussed that the surface of PPLN is screened via short-range adsorption of ambient humidity molecules [29], and the rise in temperature results in evaporation of water and local decrease in the humidity, which significantly suppress the CGM signal. Furthermore, when the relative ambient humidity is less than 30%, no CGM signal is observed in our experiments.

The dependence of CGM current signal on scan speed and temperature are clearly revealed by PCA, consistent with the proposed imaging mechanisms based on scraping and refilling of charges within domains, and changes of polarizations at domain wall. With such an understanding on CGM current, we can estimate the spontaneous polarization $P_s$ as well as the surface screening charge density $|\sigma_d|$ based on the current measured in CGM. We calculate the collected charges over the domain walls and over domains, separately, as explained in detail in SI section 2.1 and shown in Fig. S3. Passing from a downward to upward domain with the absolute polarization difference of $2P_s$, the maximum collected charges on domain walls $|Q_w|$ can be correlated with the sample polarization as $|Q_w| = 2P_s \times A_{tip}$, where $A_{tip}$ is the tip-sample contact area. The estimated value of $P_s$ is $46 \frac{\mu C}{cm^2}$ (refer to SI section 2.2 and Fig. S4), comparable to previously measured value of $P_s = 80 \pm 5 \frac{\mu C}{cm^2}$ for LiNbO$_3$ crystals [2]. Moreover, the surface charge density $|\sigma_d|$ is estimated around $2.84 \frac{\mu C}{cm^2}$ (for details refer to SI section 2.3 and Fig. S5). Both spontaneous polarization and surface charge density estimated from experimental data appear to be relatively low, perhaps because of the partial screening of polarization charges and incomplete scraping and refilling of surface charges under moving CGM probe.

## Concluding Remarks

In summary, we have used CGM and PCA to image ferroelectric domains of PPLN, and found that the CGM signal increases linearly with the scan speed while decreases via power law with the temperature. The observations are consistent with proposed imaging mechanisms of scraping and refilling of surface screening charge within domains, and polarization change across domain wall, enabling us to estimate the spontaneous polarization and the density of surface charges with order of magnitude agreement with literature data. The study demonstrates value of PCA, which helps us reduce noise level and enhance signal-to-noise ratio, making quantitative analysis of noise raw data possible.

## Acknowledgements

We acknowledge National Key Research and Development Program of China (2016YFA0201001), US National Science Foundation (CBET-1435968), and National Natural Science Foundation of China (11627801,11472236 and 51472037). This material is based in part upon work supported by the State of Washington through the University of Washington Clean Energy Institute.

## References

1.  Weis, R.S. and T.K. Gaylord, *Lithium-Niobate - Summary of Physical-Properties and Crystal-Structure.* Applied Physics a-Materials Science & Processing, 1985. **37**(4): p. 191-203.
2.  Gopalan, V., et al., *The role of nonstoichiometry in 180° domain switching of LiNbO 3 crystals.* Applied Physics Letters, 1998. **72**(16): p. 1981-1983.
3.  Lu, Y.Q., et al., *Electro-optic effect of periodically poled optical superlattice LiNbO3 and its applications.* Appl. Phys. Lett., 2000. **77**(23): p. 3719-3721.
4.  Myers, L.E., et al., *Quasi-phase-matched 1.064-μm-pumped optical parametric oscillator in bulk periodically poled LiNbO 3.* Optics letters, 1995. **20**(1): p. 52-54.
5.  Myers, L.E., et al., *QUASI-PHASE-MATCHED 1.064-MU-M-PUMPED OPTICAL PARAMETRIC OSCILLATOR IN BULK PERIODICALLY POLED LINBO3.* Opt. Lett., 1995. **20**(1): p. 52-54.
6.  Cho, Y., et al. *Ferroelectric Ultra High-Density Data Storage Based on Scanning Nonlinear Dielectric Microscopy.* in *Non-Volatile Memory Technology Symposium, 2006. NVMTS 2006. 7th Annual.* 2006.
7.  Hong, S. and Y. Kim, *Ferroelectric Probe Storage Devices*, in *Emerging Non-Volatile Memories.* 2014, Springer. p. 259-273.

8.      Soluch, V. and M. Lysakowska, in *IEEE Trans. Ultrason. Ferroelectr. Freq. Control*. 2005. p. 145.

9.      Hong, S., S.M. Nakhmanson, and D.D. Fong, *Screening mechanisms at polar oxide heterointerfaces*. Reports on Progress in Physics, 2016. **79**(7): p. 076501.

10.     Lei, S., et al., *Origin of piezoelectric response under a biased scanning probe microscopy tip across a 180° ferroelectric domain wall*. Physical Review B, 2012. **86**(13): p. 134115.

11.     Rodriguez, B.J., et al., *Domain growth kinetics in lithium niobate single crystals studied by piezoresponse force microscopy*. Applied Physics Letters, 2005. **86**(1): p. 012906.

12.     Gruverman, A., D. Wu, and J.F. Scott, *Piezoresponse force microscopy studies of switching behavior of ferroelectric capacitors on a 100-ns time scale*. Phys Rev Lett, 2008. **100**(9): p. 097601.

13.     Habicht, S., R.J. Nemanich, and A. Gruverman, *Physical adsorption on ferroelectric surfaces: photoinduced and thermal effects*. Nanotechnology, 2008. **19**(49): p. 495303.

14.     Proksch, R., *In-situ piezoresponse force microscopy cantilever mode shape profiling*. Journal of Applied Physics, 2015. **118**(7): p. 072011.

15.     Alexe, M. and A. Gruverman, *Nanoscale characterisation of ferroelectric materials: scanning probe microscopy approach*. 2013: Springer Science & Business Media.

16.     Wittborn, J., et al., *Nanoscale imaging of domains and domain walls in periodically poled ferroelectrics using atomic force microscopy*. Applied Physics Letters, 2002. **80**(9): p. 1622-1624.

17.     Kalinin, S.V. and D.A. Bonnell, *Local potential and polarization screening on ferroelectric surfaces*. Physical Review B, 2001. **63**(12): p. 125411.

18.     Strelcov, E., et al., *Direct Probing of Charge Injection and Polarization-Controlled Ionic Mobility on Ferroelectric LiNbO3 Surfaces*. Advanced Materials, 2014. **26**(6): p. 958-963.

19.     Bluhm, H., et al., *Imaging of domain-inverted gratings in LiNbO 3 by electrostatic force microscopy*. Applied physics letters, 1997. **71**(1): p. 146-148.

20.     Tong, S., et al., *Mechanical Removal and Rescreening of Local Screening Charges at Ferroelectric Surfaces*. Physical Review Applied, 2015. **3**(1): p. 014003.

21.     Hong, S., et al., *Charge gradient microscopy*. Proc Natl Acad Sci U S A, 2014. **111**(18): p. 6566-9.

22.     Hong, S., *Nanoscale phenomena in ferroelectric thin films*. 2004: Springer Science & Business Media.

23.     Li, L.L., et al., *Direct Imaging of the Relaxation of Individual Ferroelectric Interfaces in a Tensile-Strained Film*. Advanced Electronic Materials, 2017. **3**(4).

24.     Chen, Q.N., et al., *Mechanisms of electromechanical coupling in strain based scanning probe microscopy*. Applied Physics Letters, 2014. **104**(24): p. 242907.

25.     Strelcov, E., et al., *Role of measurement voltage on hysteresis loop shape in Piezoresponse Force Microscopy*. Applied Physics Letters, 2012. **101**(19): p. 192902.

26.     Seol, D., et al., *Determination of ferroelectric contributions to electromechanical response by frequency dependent piezoresponse force microscopy*. Sci Rep, 2016. **6**: p. 30579.

27.     Li, J., et al., *Strain-based scanning probe microscopies for functional materials, biological structures, and electrochemical systems*. Journal of Materiomics, 2015. **1**(1): p. 3-21.

28. Choi, Y.Y., et al., *Charge collection kinetics on ferroelectric polymer surface using charge gradient microscopy.* Scientific Reports, 2016. **6**.

29. Tong, S., et al., *Imaging Ferroelectric Domains and Domain Walls Using Charge Gradient Microscopy: Role of Screening Charges.* Acs Nano, 2016. **10**(2): p. 2568-2574.

30. Pearson, K., *LIII. On lines and planes of closest fit to systems of points in space.* The London, Edinburgh, and Dublin Philosophical Magazine and Journal of Science, 1901. **2**(11): p. 559-572.

31. Hotelling, H., *Analysis of a complex of statistical variables into principal components.* Journal of educational psychology, 1933. **24**(6): p. 417.

32. Jesse, S. and S.V. Kalinin, *Principal component and spatial correlation analysis of spectroscopic-imaging data in scanning probe microscopy.* Nanotechnology, 2009. **20**(8): p. 085714.

33. Jesse, S., B. Mirman, and S.V. Kalinin, *Resonance enhancement in piezoresponse force microscopy: Mapping electromechanical activity, contact stiffness, and Q factor.* Applied Physics Letters, 2006. **89**(2): p. 022906.

34. Bo, H.F., et al., *Drive frequency dependent phase imaging in piezoresponse force microscopy.* Journal of Applied Physics, 2010. **108**(4): p. 042003.

35. Harnagea, C., et al., *Quantitative ferroelectric characterization of single submicron grains in Bi-layered perovskite thin films.* Applied Physics a-Materials Science & Processing, 2000. **70**(3): p. 261-267.

36. Chen, Q.N., S.B. Adler, and J.Y. Li, *Imaging space charge regions in Sm-doped ceria using electrochemical strain microscopy.* Applied Physics Letters, 2014. **105**(20): p. 201602.

37. Schroder, M., et al., *Conducting Domain Walls in Lithium Niobate Single Crystals.* Advanced Functional Materials, 2012. **22**(18): p. 3936-3944.

38. Green, A.A., et al., *A Transformation for Ordering Multispectral Data in Terms of Image Quality with Implications for Noise Removal.* Ieee Transactions on Geoscience and Remote Sensing, 1988. **26**(1): p. 65-74.

# Imaging Ferroelectric Domain via Charge Gradient Microscopy Enhanced by Principal Component Analysis

## Supplementary Information

## 1.    Principal component analysis

Principal component analysis (PCA) has been performed on a set of 40 CGM images with various probe speeds, and on a separate set of 20 CGM images at various temperatures. In the analysis, we imported the images from IGOR Pro (Asylum Research AFM software) to MATLAB®, specifically the topography data and the trace and retrace current data. The data of each image was reshaped to a vector, with each vector centered around zero based on its mean value, and the matrix of experimental data **I** is constructed. The effect of drift in the series of CGM images was compensated by taking the topography of the first image as a benchmark and performing a 2D-cross-correlation (MATLAB's `xcor2` function) between topography of each image and the benchmark.

The PCA analysis was performed using MATLAB's `pca` function. The function `pca(I)` returns the principal component variances `latent` (referred here as eigenvalues, Fig. 3&4(d)), principal component `scores` (referred here as matrix of eigenvectors, Fig. 3&4(e)), and the principal component `scores` (referred here as matrix of PCA component images, Fig. 3&4(f)). The first four PCA component images for the speed-dependent and temperature-dependent CGM are shown in Figs. S1&2, respectively, where it is evident that only the first mode of PCA contains meaningful information and the higher modes mostly contain the background noises. An example of post-processing code with a typical data can be found in the following link here.

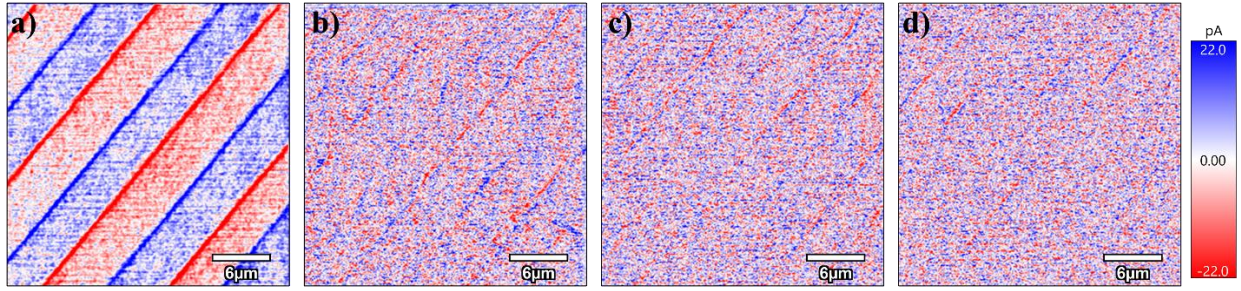

**Fig. S1** The first four PCA component images of speed-dependent data, wherein it is observed that the first component image of PCA contains the most information.

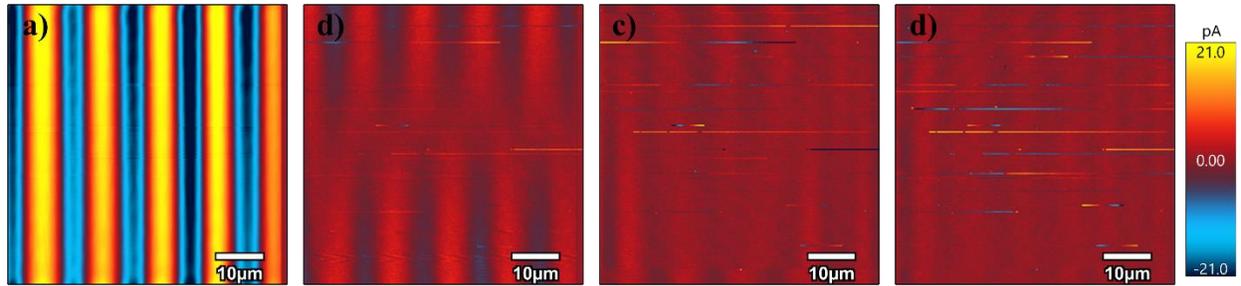

**Fig. S2** The first four PCA component images of temperature-dependent data (a-d), wherein it is observed that the first component image of PCA contains the most information.

## 2. Quantitative analysis of spontaneous polarization and surface charge density

We estimate the spontaneous polarization $P_s$ and the surface screen charge density $|\sigma_d|$ through calculating the absolute scraped charge on domains and domain walls based on CGM signals. The charge can be calculated by integrating the current with respect to time in each line scan of CGM. The estimated displacement charge associated with passing over a domain wall can be correlated with the spontaneous polarization as: $|Q_w| = 2P_s A_{tip}$, where $A_{tip}$ is the circular contact area of tip-sample junction with tip radius of 45 nm. Similarly, the screening surface charge density $|\sigma_d|$ can be estimated by averaging the collected absolute domain charges $|Q_d|$ over the number of pixels $n$ and the tip area as $|\sigma_d| = \frac{|Q_d|}{n \times A_{tip}}$, as explained in detail in the next section, and the extent of surface screening is assessed by finding the ratio of $\frac{|\sigma_d|}{P_s}$.

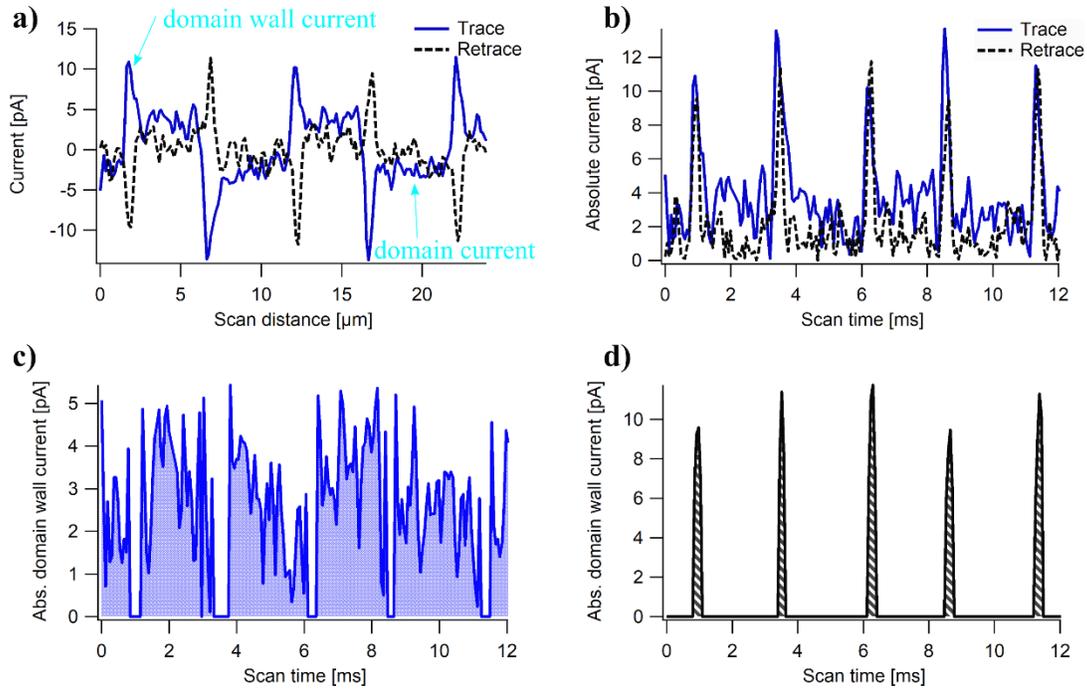

**Fig. S3** Calculation of collected domain and domain wall charges via CGM current signals. Line profiles of (a) trace and retrace CGM image as a function of scan distance; (b) absolute trace and retrace line profiles of CGM image as a function of scan time; (c) masked trace signal only over domains; and (d) masked retrace signal only over the domain walls.

### 2.1. *Calculation of collected charges over domains and domain walls*

A simple example of this analysis is visualized in Fig. S3, where a trace and a retrace line scans of a CGM image are shown in Fig. S3(a) as a function of scan distance. With the CGM probe speed given, the absolute value of these signals as a function scan time were calculated and shown in Fig. S3(b). The area under these curve is the collected charges, which were scraped over domains and domain walls. To separate these two mechanisms of charge scraping, the signals were masked based on a predefined threshold. The corresponding domain signals are obtained by masking the trace signal and removing the values bigger than the threshold of 7 pA, while the domain wall signals were acquired by ignoring the retrace values less than of 5.5 pA. The shaded area under the curve in Fig. S3(c) is equal to the absolute scraped charge over 5 domains while the shaded area in Fig. S3(d) is equal to the scraped charge over 5 domain walls.

### 2.2. *Estimation of spontaneous polarization based on domain wall charge*

The retrace signal is mostly dominated by the domain wall charges as shown in Fig. S4(a). The image was masked by replacing all the absolute values less than 5.5 pA with zero, which is shown in Fig. S4(b). The area under each line scan of this image is found with respect to the time of scan, summed over all line scans, and the total collected charge over all domain walls found to be $|Q_w| \cong 6.12$ pC. It is found that the number of current spikes corresponding to the domain wall crossing is around $n = 1040$. The estimated polarization is defined as $P_s \approx \frac{|Q_w|}{2A_{tip} \times n} = 46.22 \frac{\mu C}{cm^2}$.

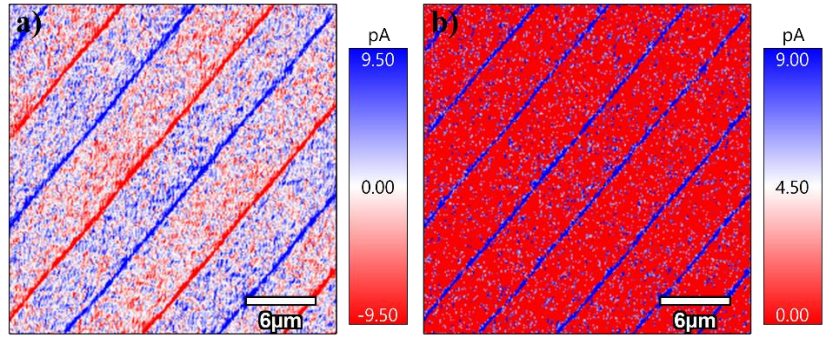

**Fig. S4** CGM retrace image (a), and the masked retrace image with threshold of 5.5 pA (b).

### 2.3. *Estimation of surface charge density based on domain charge*

As discussed in the manuscript, the trace signal contains the scraped charge over the domains. To remove the domain wall current spikes, the trace CGM image (shown in Fig. S5(a)) is masked by considering only absolute current values less than the threshold of 7 pA. Fig. S5(b) shows the absolute current values only over the domains. The sum of areas under each line scan of this image with respect to the time is the total charge collected over the domains and is equal to $|Q_d| = 9.076$ pC. The number of pixels with non-zero values (corresponding to domain charges) in Fig. S5(b) is equal to 50173 and the average collected domain charge per pixel is equal to $|Q_{dp}| = 0.1809 \frac{fC}{pixel}$. The surface charge density $|\sigma_d|$ is then estimated as $|\sigma_d| = \frac{|Q_{dp}|}{A_{tip}} 2.84 \frac{\mu C}{cm^2}$. We estimate the extent of surface screening by comparing the estimated domain charge density

$|\sigma_d|$ with the ideal surface charge of ferroelectric surface, which should be equal to polarization charges $P_s = 80 \frac{\mu C}{cm^2}$), and found 3.55% screening ratio.

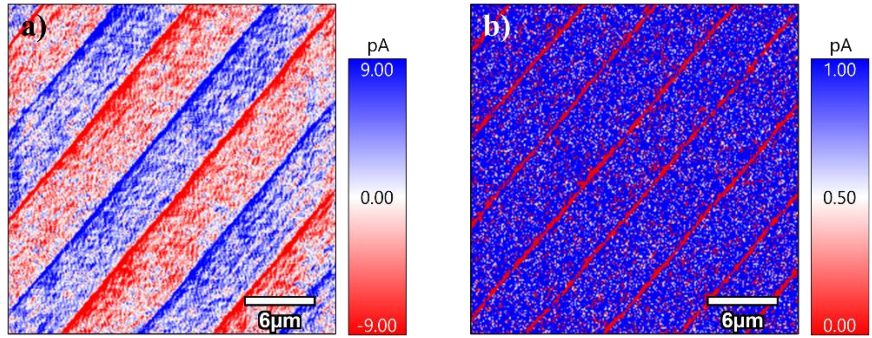

**Fig. S5** CGM trace image (a), and masked image with threshold of 7 pA (b). The masked image only captures the current signal over the domains.